\documentstyle[12pt,axodraw]{article}
\begin{document}
\baselineskip 0.6cm

\newcommand{\gsim}{\, \mathop{}_{\textstyle \sim}^{\textstyle >} \,}
\newcommand{\lsim}{\, \mathop{}_{\textstyle \sim}^{\textstyle <} \,}
\newcommand{\vev}[1]{ \langle {#1} \rangle }

\begin{titlepage}

\begin{flushright}
UCB-PTH-02/29 \\
LBNL-50970 \\
\end{flushright}

\vskip 0.5cm

\begin{center}
{\Large \bf $SO(10)$ and $SU(6)$ Unified Theories \\ 
  on an Elongated Rectangle}

\vskip 0.6cm

{\large
Lawrence J.~Hall and Yasunori Nomura
}

\vskip 0.3cm

{\it Department of Physics, University of California,
                Berkeley, CA 94720, USA}\\
{\it Theoretical Physics Group, Lawrence Berkeley National Laboratory,
                Berkeley, CA 94720, USA}

\vskip 0.6cm

\abstract{Maximally supersymmetric $SO(10)$ and $SU(6)$ unified 
theories are constructed on the orbifold $T^2/(Z_2 \times Z'_2)$, 
with one length scale $R_5$ taken much larger than the other, $R_6$. 
The effective theory below $1/R_6$ is found to be the highly 
successful $SU(5)$ theory in 5D with natural doublet-triplet 
splitting, no proton decay from operators of dimension four or 
five, unified mass relations for heavier generations only, and a 
precise prediction for gauge coupling unification.  A more unified 
gauge symmetry, and the possibility of Higgs doublets being components 
of the higher dimensional gauge multiplet, are therefore compatible 
with a large energy interval where physics is described by $SU(5)$ 
gauge symmetry in 5D. This leads to the distinctive branching ratios 
for proton decay from $SU(5)$ gauge boson exchange, $p \rightarrow 
l^+ \pi^0, l^+ K^0, \bar{\nu} \pi^+, \bar{\nu} K^+$ ($l = e, \mu$), 
for well-motivated locations for matter. Several phenomenological 
features of the higher unified gauge symmetry are discussed, including 
the role of an extra $U(1)$ gauge symmetry, which survives 
compactification, in the generation of neutrino masses.}

\end{center}
\end{titlepage}

\section{Introduction}
\label{sec:intro}

The unification of the three standard model gauge couplings with weak 
scale supersymmetry suggests a threshold for some unified physics 
at very high energies.  In previous papers we have shown that gauge 
coupling unification may occur in higher dimensional unified theories 
when the gauge symmetry is broken by boundary conditions 
\cite{Hall:2001pg, Hall:2001xb, Hall:2002ci}. The resulting explicit 
local breaking of the gauge symmetry at boundaries of the space 
does not destroy gauge coupling unification providing the volume 
of the bulk is large \cite{Hall:2001pg}. In particular, we have 
found that the simplest such theory --- $SU(5)$ in 5D --- possesses 
a set of remarkable features, making it extremely attractive 
as the effective field theory description of nature above the 
compactification scale, $M_5 = 1/R_5 \approx 10^{15}~{\rm GeV}$, 
right up to the scale of strong coupling, $M_s \approx 
10^{17}~{\rm GeV}$ \cite{Hall:2001xb, Hall:2002ci}. In this paper 
we go beyond this effective field theory, taking a closer look at 
the energy interval just below strong coupling. In particular we find 
that the positive features of the 5D effective theory are maintained 
even if a sixth dimension opens up just before strong coupling. These 
features are then seen to arise from a more symmetrical field theory, 
with gauge group $SO(10)$ \cite{SO10} or $SU(6)$ and $N=4$ supersymmetry 
from the 4D viewpoint. 

\begin{figure}
\begin{center}
    \input{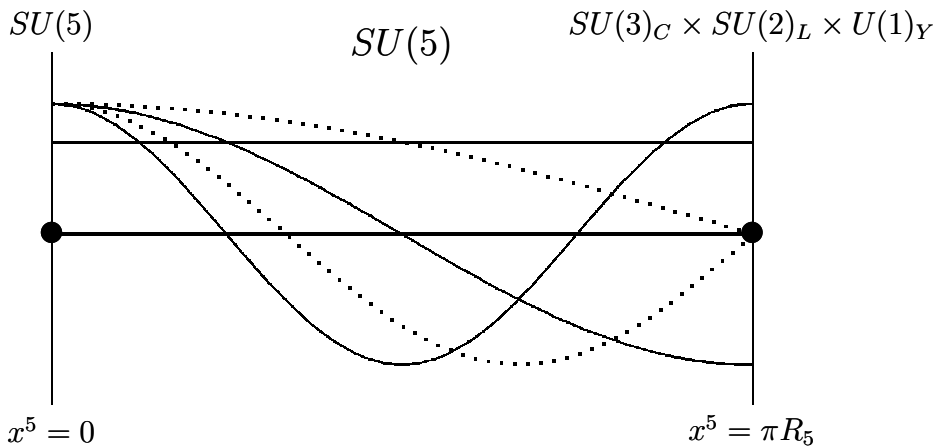}
\caption{In the fifth dimension, space is a line segment bounded by 
 branes at $x^5=0$ and at $x^5=\pi R_5$.  Here, solid and dotted lines 
 represent the profiles of gauge transformation parameters for 
 $SU(3)_C \times SU(2)_L \times U(1)_Y$, $\xi_{321}$, and 
 $SU(5)/(SU(3)_C \times SU(2)_L \times U(1)_Y)$, $\xi_X$, respectively. 
 Because $\xi_X(x^5 = \pi R_5) = 0$, explicit point defect symmetry 
 breaking occurs at the $x^5 = \pi R_5$ brane, which only respects 
 $SU(3)_C \times SU(2)_L \times U(1)_Y$ gauge symmetry.}
\label{fig:orbifold}
\end{center}
\end{figure}

The gauge symmetry of the $SU(5)$ effective theory is illustrated 
in Fig.~\ref{fig:orbifold}; it results from imposing a translation 
boundary condition under $x^5 \rightarrow x^5 + 2 \pi R_5$ of 
$(+,+,+,-,-)$ in the $SU(5)$ space. Higgs hypermultiplets in the 
${\bf 5}+\bar{\bf 5}$ ($H+\bar{H}$) representation are located in 
the bulk, while matter in $\bar{\bf 5}$ ($F$) and ${\bf 10}$ ($T$) 
representations can reside either in the bulk or on the $SU(5)$ 
invariant fixed point. No larger multiplets are needed.

Above the compactification scale the gauge couplings receive power 
law corrections, but these corrections are universal because of the 
bulk $SU(5)$ gauge symmetry. However, the $SU(5)$ breaking defect at 
$x^5 = \pi R_5$ induces a relative logarithmic running of the gauge 
couplings above $M_5$, which follows from the pattern of $SU(5)$ 
breaking in the Kaluza-Klein towers of the gauge and Higgs 
supermultiplets. The resulting correction to gauge coupling unification
\begin{equation}
  \delta\alpha_s \simeq -\frac{3}{7\pi} \alpha_s^2 \ln\frac{\pi M_s}{M_5},
\label{eq:as-formula}
\end{equation}
precisely corrects the central value for the prediction for gauge
coupling unification from $\alpha_s(M_Z) \simeq 0.130$ to 
$\simeq 0.118$ \cite{Hall:2001xb}, which should be compared with the 
experimental value of $\alpha_s^{\rm exp}(M_Z) = 0.117 \pm 0.002$ 
\cite{Groom:2000in}.  If the third generation $T_3 + F_3$ is located 
at $x^5 = 0$, a unified mass relation occurs for $m_b/m_\tau$. This 
unified mass relation holds at the compactification scale $M_5$, which 
is smaller than the 4D unification scale, so that the 4D prediction 
for the bottom quark mass is corrected by
\begin{equation}
  \frac{\delta m_b}{m_b} 
    \simeq - \frac{20g^2 - 5y_t^2}{112\pi^2} \ln\frac{\pi M_s}{M_5},
\label{eq:delta-mb-Mc}
\end{equation}
serving to slightly improve the agreement with data \cite{Hall:2002ci}.

Breaking $SU(5)$ by the translation boundary condition is not just a
simple way to break the gauge symmetry: it also leads to a very
elegant understanding of why there is a light Higgs doublet but not a
light colored triplet \cite{Kawamura:2001ev}. Furthermore the theory 
possesses a $U(1)_R$ symmetry which leads to the removal of all baryon 
number violating operators at dimension four and five, solving yet 
another long standing difficulty of 4D supersymmetric grand unified 
theories \cite{Hall:2001pg}. Finally, matter which resides on the 
$SU(5)$ fixed point is expected to be heavy, since its mass is not 
suppressed by a volume dilution factor, and to exhibit $SU(5)$ mass 
relations, reflecting the symmetry at that point \cite{Hall:2001pg}. 
On the other hand, matter in the bulk will be light, and will not 
respect $SU(5)$ mass relations because the zero mode structure is 
greatly affected by the $SU(5)$ breaking defect at $x^5= \pi R_5$ 
\cite{Hall:2001pg, Hebecker:2001wq} --- hence there is a successful 
correlation: only the heavier fermions are expected to exhibit unified 
mass relations \cite{Hall:2001xb, Hall:2001zb, Hall:2001rz}.
A simple, realistic grand unified construction is complete.

While there are certainly other issues that can be addressed within 
the context of the 5D effective theory, such as the location of each 
matter field and the breaking of supersymmetry, in this paper we study
how this effective theory can emerge from a more unified theory at
higher energies. The 5D theory may emerge directly from string theory 
at $M_s$, or there may be some strongly coupled field theory at $M_s$. 
In this paper we explore the possibility that the minimal 5D $SU(5)$ 
theory of Ref.~\cite{Hall:2001xb} is the valid effective field theory 
over a large energy interval from $M_5$ to $M_6 = 1/R_6$, but is 
incorporated into a 6D field theory at $M_6$ which is still perturbative. 
We will show that the successes of the 5D theory are maintained as 
long as $M_6/M_5$ is large enough, but this still allows a sufficiently 
large energy interval $M_s/M_6$ to study the 6D field theory.  
In constructing a complete $SU(5)$ theory in 5D, an additional $U(1)$ 
gauge interaction was needed for a variety of reasons --- in particular 
to understand the see-saw mechanism for neutrino masses \cite{Seesaw}. 
In the 6D context this $U(1)$ arises naturally if the rank of the bulk 
gauge group $G$ is one larger than that of $SU(5)$. We are led to study 
two cases for $G$: $SO(10)$ and $SU(6)$. In both cases we take the bulk 
to contain a maximal amount of supersymmetry, $N=2$ in 6D and therefore 
$N=4$ in 4D, which guarantees that the theory is free from all anomalies 
providing the 4D anomalies vanish. In the case of $G = SU(6)$, gauge 
and Higgs fields may be unified into a single 6D gauge supermultiplet.

\section{Theories with $N=2$ Supersymmetry in 6D}
\label{sec:theory}

In this section we construct a set of 6D supersymmetric unified 
theories, which provide effective descriptions of nature just below 
the cutoff scale $M_s$ where the theories become strongly coupled 
and embedded into some more fundamental theory.  Below the scale of 
$M_6 \approx 10^{16}-10^{17}~{\rm GeV}$, which is taken to be a factor 
of a few smaller than $M_s$, these theories are reduced to the 5D 
$SU(5)$ theory of Ref.~\cite{Hall:2001xb} (with an extra $U(1)$ factor) 
which is an appropriate effective field theory describing the physics 
over a wide energy interval from $M_6$ down to the scale of the fifth 
dimension, $M_5 \approx 10^{15}~{\rm GeV}$.

In general, 6D supersymmetric gauge theories compactified on orbifolds 
are subject to stringent constraints from anomaly cancellation: 
not only low energy 4D anomalies arising on fixed points but also 
anomalies in the 6D bulk must be canceled \cite{Hebecker:2001jb}. 
In theories with 6D $N=1$ supersymmetry, these constraints are 
extremely restrictive, making it difficult to find completely 
realistic anomaly-free theories.  Although it is possible to 
construct such theories, in most cases we have to rely on the 
Green-Schwarz mechanism to cancel the bulk anomalies, requiring extra 
axion-like states in the low-energy theory.  Therefore, in this paper 
we consider 6D $N=2$ theories in which the cancellation of the bulk 
anomalies is automatic due to the vector-like nature of these theories 
in 6D.  We require that our theories have two separate mass scales $M_5$ 
and $M_6 (\gg M_5)$ and that they reduce to 5D $N=1$ theories between 
$M_5$ and $M_6$ and 4D $N=1$ theories below $M_5$.  Then we find 
that $T^2/(Z_2 \times Z'_2)$ is the unique simple orbifold on which 
our 6D theories are compactified --- we are led to consider 6D $N=2$ 
supersymmetric gauge theories with gauge group $G$, compactified on 
the $T^2/(Z_2 \times Z'_2)$ orbifold having two different radii 
$R_5 \gg R_6$.  The structure of these theories are very rich, having 
four 5D fixed lines and four 4D fixed points with differing gauge and 
supersymmetries.  While 6D $N=2$ supersymmetry allows only the gauge 
multiplet to be located in the bulk, we can introduce a variety of 
matter and Higgs fields on 4D or 5D fixed sub-spaces.  In sub-section 
\ref{subsec:higgs-hyper} we construct completely realistic theories 
based on $G=SO(10)$ and $G=SU(6)$, in which the Higgs fields are 
located on a 5D fixed line. In sub-section \ref{subsec:higgs-gauge} 
we consider the theories where the Higgs fields are unified with the 
gauge fields into a single 6D gauge supermultiplet. This gauge-Higgs 
unification selects the gauge group $G = SU(6)$.

\subsection{Gauge unification on asymmetric $T^2/(Z_2 \times Z'_2)$}
\label{subsec:higgs-hyper}

The compactification on the $T^2/(Z_2 \times Z'_2)$ orbifold is obtained 
by identifying points of the infinite plane $R^2$ under four operations, 
${\cal Z}_5: (x^5,x^6) \rightarrow (-x^5,x^6)$, 
${\cal Z}_6: (x^5,x^6) \rightarrow (x^5,-x^6)$, 
${\cal T}_5: (x^5,x^6) \rightarrow (x^5+2\pi R_5,x^6)$ and 
${\cal T}_6: (x^5,x^6) \rightarrow (x^5,x^6+2\pi R_6)$.  Here, for
simplicity, we have taken the two translations ${\cal T}_5$ and 
${\cal T}_6$ to be in orthogonal directions.  We take two radii to be 
highly asymmetric, $R_5 \gg R_6$, as discussed before. In 6D $N=2$ 
theories, the only field which can be introduced in the 6D bulk is 
a gauge supermultiplet.  Under the 4D $N=1$ superfield language, 
this multiplet is decomposed into a vector superfield $V$ and three 
chiral superfields $\Sigma_5$, $\Sigma_6$ and $\Phi$, where $\Sigma_5$ 
($\Sigma_6$) contains the fifth (sixth) component of the gauge 
field, $A_5$ ($A_6$), in its lowest component; all these superfields 
transform as adjoint under the gauge group $G$.

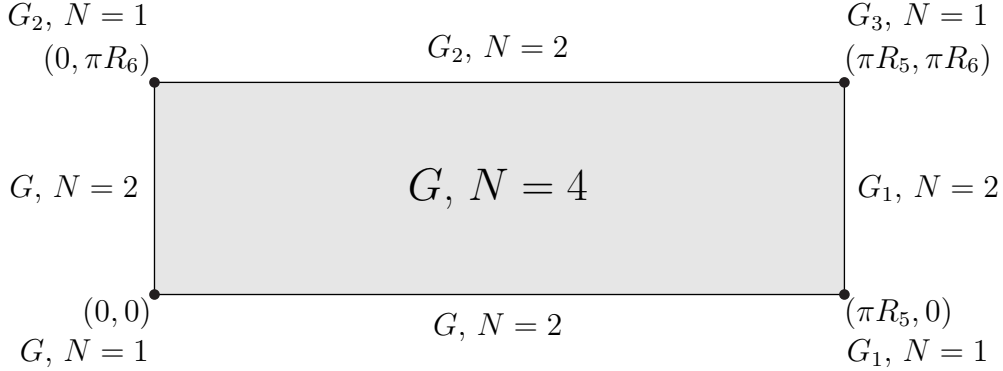
\begin{figure}
\begin{center}
\begin{picture}(300,130)(0,-5)
  \GBox(20,20)(280,100){0.9}
  \Text(150,60)[c]{\Large $G$, $N=4$}
  \Vertex(20,20){2}   \Text(19,19)[tr]{$(0,0)$} 
    \Text(18,3)[tr]{$G$, $N=1$}
  \Vertex(280,20){2}  \Text(281,19)[tl]{$(\pi R_5,0)$} 
    \Text(282,3)[tl]{$G_1$, $N=1$}
  \Vertex(20,100){2}  \Text(19,106)[br]{$(0,\pi R_6)$} 
    \Text(18,122)[br]{$G_2$, $N=1$}
  \Vertex(280,100){2} \Text(281,106)[bl]{$(\pi R_5,\pi R_6)$} 
    \Text(282,122)[bl]{$G_3$, $N=1$}
  \Text(14,60)[r]{$G$, $N=2$}  \Text(286,60)[l]{$G_1$, $N=2$}
  \Text(150,13)[t]{$G$, $N=2$} \Text(150,110)[b]{$G_2$, $N=2$}
\end{picture}
\caption{In the 2D bulk, $(x^5,x^6)$, the physical space is a rectangle 
with four sides and four corners. Each side and corner has its own 
gauge symmetry and number of supersymmetries, which are shown in the 
figure.  Here, $G_1$, $G_2$ and $G_3$ are subgroups of $G$, and the 
number of supersymmetries is that of the 4D picture: 6D $N=2$ is written 
as $N=4$ (in 4D), and 5D $N=1$ as $N=2$ (in 4D).}
\label{fig:higgs-line}
\end{center}
\end{figure}

The boundary conditions for the gauge multiplet are given by 
\begin{equation}
  \begin{array}{ccccc}
    V(x^5,x^6)        &=& V(-x^5,x^6)         &=& V(x^5,-x^6),         \\
    \Sigma_5(x^5,x^6) &=& -\Sigma_5(-x^5,x^6) &=& \Sigma_5(x^5,-x^6),  \\
    \Sigma_6(x^5,x^6) &=& \Sigma_6(-x^5,x^6)  &=& -\Sigma_6(x^5,-x^6), \\
    \Phi(x^5,x^6)     &=& -\Phi(-x^5,x^6)     &=& -\Phi(x^5,-x^6),
  \end{array}
\label{eq:bc-Z}
\end{equation}
and
\begin{equation}
  \begin{array}{ccccc}
    V(x^5,x^6)        &=& P_5\, V(x^5+2\pi R_5,x^6)\, P_5^{-1} 
      &=& P_6\, V(x^5,x^6+2\pi R_6)\, P_6^{-1}, \\
    \Sigma_5(x^5,x^6) &=& P_5\, \Sigma_5(x^5+2\pi R_5,x^6)\, P_5^{-1} 
      &=& P_6\, \Sigma_5(x^5,x^6+2\pi R_6)\, P_6^{-1}, \\
    \Sigma_6(x^5,x^6) &=& P_5\, \Sigma_6(x^5+2\pi R_5,x^6)\, P_5^{-1} 
      &=& P_6\, \Sigma_6(x^5,x^6+2\pi R_6)\, P_6^{-1}, \\
    \Phi(x^5,x^6)     &=& P_5\, \Phi(x^5+2\pi R_5,x^6)\, P_5^{-1} 
      &=& P_6\, \Phi(x^5,x^6+2\pi R_6)\, P_6^{-1},
  \end{array}
\label{eq:bc-T}
\end{equation}
where $P_5$ and $P_6$ are matrices acting on the gauge space, which 
in general do not commute with the gauge generators.  By choosing 
these matrices, we can have a variety of patterns for the gauge 
breaking structure in the extra dimensions.  The resulting gauge and 
supersymmetry structure in the 2D extra dimensions is summarized in 
Fig.~\ref{fig:higgs-line}. The gauge and supersymmetries at each fixed 
point are given by the intersection of those on the adjacent fixed 
lines; for example, $G_3 = G_1 \cap G_2$.

To construct theories which reduce below $M_6$ to the 5D $SU(5)$ 
theory with an extra $U(1)$ gauge interaction, we take either 
$G=SO(10)$ or $G=SU(6)$.  We first consider the case of $G=SO(10)$. 
The $SO(10)$ unified theories in 6D were first considered in 
Refs.~\cite{Asaka:2001eh, Hall:2001xr}. In particular, 6D $SO(10)$ 
theories on $T^2/(Z_2 \times Z'_2)$ have been constructed in 
Ref.~\cite{Hall:2001xr}, and here we follow the notation used there.
For alternative implementations of $SO(10)$ in higher dimensions, 
see Ref.~\cite{Li:2001wz}. The generators $T^a$ of $SO(10)$ 
are imaginary and antisymmetric $10 \times 10$ matrices, which are 
conveniently written as tensor products of $2 \times 2$ and 
$5 \times 5$ matrices, giving $\sigma_0 \otimes A_5$, $\sigma_1 
\otimes A_5$, $\sigma_2 \otimes S_5$ and $\sigma_3 \otimes A_5$ as 
a complete set.  Here $\sigma_0$ is the $2 \times 2$ unit matrix and 
$\sigma_{1,2,3}$ are the Pauli spin matrices; $S_5$ and $A_5$ are 
$5 \times 5$ matrices that are real and symmetric, and imaginary and 
antisymmetric, respectively.  The $\sigma_0 \otimes A_5$ and 
$\sigma_2 \otimes S_5$ generators form an $SU(5) \otimes U(1)_X$ 
subgroup of $SO(10)$, with $U(1)_X$ given by $\sigma_2 \otimes I_5$. 
We choose our basis so that the standard model gauge group is contained 
in this $SU(5)$ (Georgi-Glashow $SU(5)$ \cite{Georgi:1974sy}), 
with $SU(3)_C$ contained in $\sigma_0 \otimes A_3$ and $\sigma_2 
\otimes S_3$ and $SU(2)_L$ contained in $\sigma_0 \otimes A_2$ 
and $\sigma_2 \otimes S_2$, where $A_3$ and $S_3$ have indices 1,2,3 
and $A_2$ and $S_2$ have indices 4,5. 

In order to obtain an effective 5D $SU(5)$ theory below $M_6 = 1/R_6$, 
we have to choose $P_6 = \sigma_2 \otimes I_5$, giving 
$G_2 = SU(5) \times U(1)_X$.  For $P_5$, we have two choices 
$P_5 = \sigma_0 \otimes {\rm diag}(1,1,1,-1,-1)$ and 
$P_5 = \sigma_2 \otimes {\rm diag}(1,1,1,-1,-1)$, giving 
$G_1 = SU(4)_C \times SU(2)_L \times SU(2)_R$ (Pati-Salam group 
\cite{Pati:1974yy}) and $G_1 = SU(5)' \times U(1)'_X$ (flipped 
$SU(5)$ \cite{Barr:1982qv}), respectively.  In either case, 
$G_3 = SU(3)_C \times SU(2)_L \times U(1)_Y \times U(1)_X$, and the 
unbroken gauge symmetries below $M_5 = 1/R_5$ is the standard model 
gauge group with an extra $U(1)_X$ \cite{Hall:2001xr, Asaka:2001eh}.
The massless fields arising from the gauge multiplet are only vector 
superfields, $V$, of the $SU(3)_C \times SU(2)_L \times U(1)_Y \times 
U(1)_X$ gauge group; all the other fields are heavy with masses larger 
than $\sim M_5$.  Below, we construct theories concentrating on the 
case with $G_1 = SU(4)_C \times SU(2)_L \times SU(2)_R$, but 
completely realistic theories are also obtained in the other choice 
of $G_1 = SU(5)' \times U(1)'_X$.

Having fixed the gauge symmetry structure, we now consider the Higgs 
fields.  In the effective 5D $SU(5)$ theory below $M_6$, the standard 
model Higgs doublets arise from two hypermultiplets of the 
${\bf 5} + \bar{\bf 5}$ representation, $\{ H, H^c \} 
+ \{ \bar{H}, \bar{H}^c \}$, located in the bulk.  Here, we have used 
4D $N=1$ superfield language: $H$ and $\bar{H}^c$ ($H^c$ and $\bar{H}$) 
are 4D chiral superfields transforming as ${\bf 5}$ ($\bar{\bf 5}$) 
under $SU(5)$.  This implies that we have to introduce Higgs 
hypermultiplets on the 5D fixed line, either $x^6=0$ or $x^6=\pi R_6$, 
in our 6D theory.  Although we can construct realistic theories 
in both cases, here we choose to put them on the $x^6=\pi R_6$ 
fixed line with quantum numbers given by $\{ H, H^c \}({\bf 5}, -2)$ 
and $\{ \bar{H}, \bar{H}^c \}(\bar{\bf 5}, 2)$, where the numbers in 
parentheses represent gauge quantum numbers for unconjugated chiral 
superfields under the $SU(5) \times U(1)_X$ gauge group, which is 
unbroken on the $x^6=\pi R_6$ fixed line.\footnote{
We have normalized $U(1)_X$ charges such that ${\bf 10}$ of $SO(10)$ 
decomposes into $({\bf 5}, -2) + (\bar{\bf 5}, 2)$ under the 
$SU(5) \times U(1)_X$ subgroup.}
(In the case of the Higgs on the $x^6=0$ fixed line, it arises from 
a single hypermultiplet $\{ H, H^c \}$, transforming as ${\bf 10}$ 
under $SO(10)$.)  In general, the boundary conditions for a 
hypermultiplet $\{ \Phi, \Phi^c \}$ located on a fixed line with 
a constant $x^6$ are given by 
\begin{equation}
  \begin{array}{ccccc}
    \Phi(x^5)   &=& \Phi(-x^5)    
      &=& \eta_{\Phi} P_5 \cdot \Phi(x^5+2\pi R_5), \\
    \Phi^c(x^5) &=& -\Phi^c(-x^5) 
      &=& \eta_{\Phi} P_5 \cdot \Phi^c(x^5+2\pi R_5),
  \end{array}
\label{eq:bc-hyper-5}
\end{equation}
where $\eta_{\Phi} = \pm 1$, and the matrix $P_5$ acts on the gauge 
space.  If $x^6 = \pi R_6$, we have to use $\hat{P}_5$, instead of 
$P_5$, which is obtained by projecting $P_5$ on the $SU(5) \times 
U(1)_X$ gauge space.  Choosing $\eta_H = \eta_{\bar{H}} = -1$, we 
obtain only the two Higgs doublets of the minimal supersymmetric 
standard model (MSSM) at low energies.  Thus, at this stage, the low 
energy matter content below $\sim M_5$ is the vector multiplets of 
$SU(3)_C \times SU(2)_L \times U(1)_Y \times U(1)_X$ and the two 
Higgs doublets of the MSSM.

How about quarks and leptons?  Since our theory has $M_5 \approx 
10^{15}~{\rm GeV}$, which we will see in more detail later, the first 
generation matter coming from a ${\bf 10}$ representation of $SU(5)$ 
must propagate in the fifth dimension to avoid too rapid proton 
decay caused by an exchange of the broken gauge bosons. 
This implies that we have to put two hypermultiplets 
$\{ T_1 + T_1^c \}({\bf 10}, 1) + \{ T'_1 + T'^c_1 \}({\bf 10}, 1)$ 
with $\eta_{T_1} = -\eta_{T'_1} = 1$ on the $x^6 = \pi R_6$ fixed 
line.  We then obtain MSSM quark and lepton superfields $Q_1$, $U_1$ 
and $E_1$ at low energies as zero modes of these multiplets.
For the other matter fields, we have three options: introducing on 
a 4D fixed point, on a short fixed line with a constant $x^5$, 
or on a long 5D fixed line with a constant $x^6$.  We can make a 
choice for each matter field from these options.  While there are 
many possible matter configurations leading to realistic fermion 
mass matrices, here we focus on the case where the bottom and tau 
Yukawa couplings are unified around the unified mass scale reflecting 
the underlying $SU(5)$ gauge structure \cite{Chanowitz:1977ye}. 
This forces us to introduce the third generation matter either on the 
$(x^5,x^6)=(0,\pi R_6)$ fixed point or on the $x^5=0$ fixed line, for 
the present choice of the Higgs location.  As an example, here we 
choose to put $T_3$ and $F_3$ on the fixed point and the fixed line, 
respectively: we introduce a chiral superfield $T_3({\bf 10}, 1)$ 
on the $SU(5) \times U(1)_X$ fixed point at $(x^5,x^6)=(0,\pi R_6)$, 
and a hypermultiplet $\{ \Psi_3 + \Psi_3^c \}({\bf 16})$ on the 
$SO(10)$ fixed line of $x^5=0$.  The general boundary conditions 
for a hypermultiplet $\{ \Phi, \Phi^c \}$ located on a fixed line 
with a constant $x^5$ are given by 
\begin{equation}
  \begin{array}{ccccc}
    \Phi(x^6)   &=& \Phi(-x^6)    
      &=& \zeta_{\Phi} \hat{P}_6 \cdot \Phi(x^6+2\pi R_6), \\
    \Phi^c(x^6) &=& -\Phi^c(-x^6) 
      &=& \zeta_{\Phi} \hat{P}_6 \cdot \Phi^c(x^6+2\pi R_6),
  \end{array}
\label{eq:bc-hyper-6}
\end{equation}
where $\zeta_{\Phi} = \pm 1$, and $\hat{P}_6$ is a matrix obtained 
by projecting $P_6$ on the corresponding gauge space unbroken on the 
fixed line.  Choosing $\zeta_{\Psi_3}$ appropriately, we find that 
$\bar{\bf 5} + {\bf 1}$ components of $SU(5)$ remain as zero modes from 
$\{ \Psi_3, \Psi_3^c \}$: in the effective 5D $SU(5) \times U(1)_X$ 
theory below $M_6$, the hypermultiplet $\{ \Psi_3, \Psi_3^c \}$ 
reproduces brane fields $F_3(\bar{\bf 5}, -3) + N_3({\bf 1}, 5)$ 
localized on the $SU(5) \times U(1)_X$ invariant fixed point at 
$x^5 = 0$.  Thus, together with $T_3({\bf 10}, 1)$ located on the 
$(x^5,x^6)=(0,\pi R_6)$ fixed point, we recover a complete set of the 
third generation matter $T_3+F_3+N_3$ on the $x^5 = 0$ brane in the 
effective 5D $SU(5) \times U(1)_X$ theory below $M_6$.

The configuration for the other matter fields are less restrictive.
The only significant constraint is that $T_2$ and $F_2$ cannot both 
be confined to the $x^5 = 0$ plane, to avoid the unwanted $SU(5)$ 
mass relation for $m_s/m_\mu$: at least one of $T_2$ and $F_2$ must 
propagate in the fifth dimension or be located on subspaces with 
$x^5 = \pi R_5$.  There are many possibilities which satisfy this 
criteria and lead to realistic fermion mass matrices, but here we 
do not exhaust all of these possibilities; rather we present some new 
mechanisms for understanding matter quantum numbers which can be 
implemented but cannot be fully understood in the 5D $SU(5)$ theory 
context.  First, in the present 6D theory, we can introduce matter 
fields on the $x^5 = \pi R_5$ fixed line, forming representations 
under $SU(4)_C \times SU(2)_L \times SU(2)_R$ such as 
$({\bf 4}, {\bf 2}, {\bf 1})$ or $(\bar{\bf 4}, {\bf 1}, {\bf 2})$.
In the effective 5D $SU(5)$ theory below $M_6$, these fields are 
reduced to brane fields localized on the $SU(5)$ breaking fixed point 
at $x^5 = \pi R_5$ and thus do not necessarily represent properties 
for $SU(5)$ matter; for instance, these fields are not subject to 
gauge boson mediated proton decay or $SU(5)$ Yukawa relations.
Nevertheless, the hypercharges for these fields are appropriately 
quantized, since they come from a multiplet of non-Abelian gauge group, 
$SU(4)_C \times SU(2)_L \times SU(2)_R$.  Therefore, although there 
is no reason in the effective 5D theory for why hypercharges for these 
fields are appropriately quantized (since the unbroken gauge symmetry 
on the $x^5 = \pi R_5$ brane is $SU(3)_C \times SU(2)_L \times U(1)_Y 
\times U(1)_X$), we can understand the quantization in the more 
fundamental (higher dimensional) theory. This provides a general way 
of understanding the quantization of $U(1)$ charges which are arbitrary 
in the effective field theory of interest.  For instance, we can 
imagine that $U(1)_X$ charges for matter fields introduced on the 
$SU(5) \times U(1)_X$ fixed line will be quantized in a similar way 
in some higher dimensional theory above $M_s$.\footnote{
Another intriguing mechanism for charge quantization is to use 
an anomaly inflow in the bulk through the Chern-Simons term 
\cite{Hall:2002rk}.  Suppose we introduce $T({\bf 10})$ on the 
$SU(5)$ brane at $x^5=0$ and $D(\bar{\bf 3}, {\bf 1}, \alpha/3) + 
L({\bf 1}, {\bf 2}, -\alpha/2)$ on the $SU(3)_C \times SU(2)_L \times 
U(1)_Y$ brane at $x^5=\pi R_5$, in the effective 5D $SU(5)$ theory.
This theory is consistent only if $\alpha=1$, in which case we can 
cancel all gauge anomalies by introducing the Chern-Simons term in 
the bulk with an appropriate coefficient. This mechanism even 
allows a fractional quantization of $U(1)$ charges: in a way which 
does not arise from embedding the $U(1)$ factor together with the other 
gauge factors in a larger non-Abelian gauge group \cite{Hall:2002rk}.}

We also find an interesting mechanism of realizing textures for 
fermion mass matrices.  Suppose we introduce some of the matter fields 
on the $x^5 = 0$ plane (either the fixed line or one of two fixed 
points) and some on the $x^5 = \pi R_5$ plane.  In this case, the 
fields on $x^5 = 0$ cannot couple to those on $x^5 = \pi R_5$ due to 
locality in the extra dimensions. This leads to texture zeros in 
Yukawa matrices, which are not guaranteed by any symmetry of the low 
energy effective field theory.  Clearly, this mechanism can also be 
used in the 5D $SU(5)$ theory, although in this case we have to 
assume an appropriate $U(1)_Y$ charge quantization for matter on 
the $x^5 = \pi R_5$ brane.

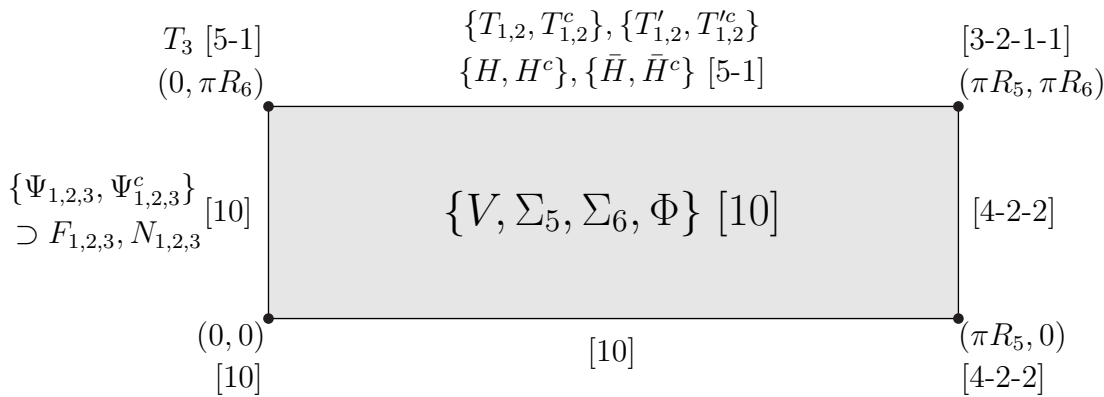
\begin{figure}
\begin{center}
\begin{picture}(300,135)(0,-5)
  \GBox(20,20)(280,100){0.9}
  \Text(150,60)[c]{\Large $\{ V, \Sigma_5, \Sigma_6, \Phi \}$ [10]}
  \Vertex(20,20){2}   \Text(19,19)[tr]{$(0,0)$} 
    \Text(18,3)[tr]{[10]}
  \Vertex(280,20){2}  \Text(281,19)[tl]{$(\pi R_5,0)$} 
    \Text(282,3)[tl]{[4-2-2]}
  \Vertex(20,100){2}  \Text(19,106)[br]{$(0,\pi R_6)$} 
    \Text(18,122)[br]{$T_3$ [5-1]}
  \Vertex(280,100){2} \Text(281,106)[bl]{$(\pi R_5,\pi R_6)$} 
    \Text(282,122)[bl]{[3-2-1-1]}
  \Text(-7,66)[br]{$\{ \Psi_{1,2,3}, \Psi^c_{1,2,3} \}$}
  \Text(-5,57)[tr]{$\supset F_{1,2,3}, N_{1,2,3}$}
  \Text(14,60)[r]{[10]} \Text(150,13)[t]{[10]}
  \Text(286,60)[l]{[4-2-2]}
  \Text(150,128)[b]{$\{ T_{1,2}, T^c_{1,2} \}, \{ T'_{1,2}, T'^c_{1,2} \}$}
  \Text(150,110)[b]{$\{ H, H^c \}, \{ \bar{H}, \bar{H}^c \}$ [5-1]}
\end{picture}
\caption{An example of the matter configuration in the 6D $SO(10)$ 
theory on asymmetric $T^2/(Z_2 \times Z'_2)$.  The numbers in the 
square brackets represent unbroken gauge symmetries on the 
corresponding (sub-)spaces: 10, 5-1, 4-2-2 and 3-2-1-1 denote 
$SO(10)$, $SU(5) \times U(1)_X$, $SU(4)_C \times SU(2)_L \times SU(2)_R$ 
and $SU(3)_C \times SU(2)_L \times U(1)_Y \times U(1)_X$, respectively.}
\label{fig:matter-config}
\end{center}
\end{figure}

While it is quite interesting to pursue completely realistic and 
predictive theories of flavor within the present framework using 
the above mechanisms, here we just present a simple example of 
matter configuration which reduces to that of Ref.~\cite{Hall:2002ci} 
in the effective 5D $SU(5) \times U(1)_X$ theory below $M_6$.  We 
introduce two hypermultiplets $\{ \Psi_1 + \Psi_1^c \}({\bf 16})$ and 
$\{ \Psi_2 + \Psi_2^c \}({\bf 16})$ on the $x^5=0$ fixed line.
Choosing $\zeta_{\Psi_1} = \zeta_{\Psi_2} = \zeta_{\Psi_3}$, these 
hypermultiplets give $F_{1,2}(\bar{\bf 5}, -3) + N_{1,2}({\bf 1}, 5)$
on the $x^5 = 0$ brane in the effective 5D theory.  The only remaining 
field is the second generation matter coming from ${\bf 10}$ of $SU(5)$, 
which we introduce on the $x^6 = \pi R_6$ brane as two hypermultiplets 
$\{ T_2 + T_2^c \}({\bf 10}, 1) + \{ T'_2 + T'^c_2 \}({\bf 10}, 1)$ 
with $\eta_{T_2} = -\eta_{T'_2} = 1$.  This completes three 
generations of matter, $T_{1,2,3}, F_{1,2,3}$ and $N_{1,2,3}$, 
for low energy fields below $M_5$.  This example of matter 
configuration is summarized in Fig.~\ref{fig:matter-config}.

The Yukawa couplings are introduced on the $(x^5,x^6)=(0,\pi R_6)$ 
fixed point.  Since the gauge symmetry on this fixed point is 
$SU(5) \times U(1)_X$, they take the form
\begin{equation}
  W = \hat{T}\hat{T}H + \hat{T}\hat{F}\bar{H} + \hat{F}\hat{N}H,
\label{eq:yukawa}
\end{equation}
where $\hat{T}$, $\hat{F}$ and $\hat{N}$ run for all the components 
of matter fields in ${\bf 10}$, $\bar{\bf 5}$ and ${\bf 1}$
representations of $SU(5)$, respectively.  For instance, in the 
example of matter configuration in Fig.~\ref{fig:matter-config}, 
$\hat{T}$ runs for $T_3, T_{1,2}$ and $T'_{1,2}$; $\hat{F}$ for 
$F_{1,2,3} \subset \Psi_{1,2,3}$; and $\hat{N}$ for $N_{1,2,3} 
\subset \Psi_{1,2,3}$.\footnote{
We could also introduce Yukawa couplings for $T_{1,2}$ and 
$T'_{1,2}$ on the $(x^5,x^6)=(\pi R_5,\pi R_6)$ fixed point, which 
do not respect the $SU(5)$ symmetry.}
Here, we have omitted coefficients for the operators suppressed by 
appropriate powers of $M_s$.  In the low energy 4D theory, these 
couplings reproduce usual Yukawa couplings of the MSSM (with neutrino 
Yukawa couplings).  Since various matter fields propagate in differing 
dimensions, various 4D Yukawa couplings have suppressions by powers of 
different volume factors. In particular, if a matter field propagates 
in the fifth (sixth) dimension, it carries a suppression factor 
$\epsilon_5 = (M'_5/M_s)^{1/2}$ ($\epsilon_6 = (M'_6/M_s)^{1/2}$), 
where $M'_5 \equiv M_5/\pi$ ($M'_6 \equiv M_6/\pi$).  In the example 
of Fig.~\ref{fig:matter-config}, this leads to 
\begin{equation}
  W_{4D} \approx 
  \pmatrix{
     T_1 & T_2 & T_3 \cr
  }
  \pmatrix{
     \epsilon_5^2 & \epsilon_5^2 & \epsilon_5      \cr
     \epsilon_5^2 & \epsilon_5^2 & \epsilon_5      \cr
     \epsilon_5   & \epsilon_5   & \underline{1}   \cr
  }
  \pmatrix{
     T_1 \cr T_2 \cr T_3 \cr
  } H
\nonumber\\
  + \epsilon_6 \pmatrix{
     T_1 & T_2 & T_3 \cr
  }
  \pmatrix{
     \epsilon_5      & \epsilon_5      & \epsilon_5      \cr
     \epsilon_5      & \epsilon_5      & \epsilon_5      \cr
     \underline{1}   & \underline{1}   & \underline{1}   \cr
  }
  \pmatrix{
     F_1 \cr F_2 \cr F_3 \cr
  } \bar{H}.
\label{eq:4D-yukawa}
\end{equation}
Here we have displayed only the gross structure that follows from the 
volume suppression factors, and omitted the coupling parameters of the 
brane-localized Yukawa interactions. Only underlined entries respect 
$SU(5)$, since the other entries involve $T_{1,2}$ which actually 
represent quarks and leptons from differing $SU(5)$ bulk multiplets. 
The only unified mass relation is for $b/\tau$.  This mass matrix 
structure is the same as that of Ref.~\cite{Hall:2002ci}, except 
that now there is an extra suppression factor $\epsilon_6$ in the 
$TF\bar{H}$ Yukawa couplings relative to the $TTH$ ones. As we will see 
later in section \ref{sec:unif}, we can imagine $\epsilon_5 \simeq 0.1$ 
and $\epsilon_6 \simeq 0.3$ as a realistic parameter region. Thus 
this extra suppression factor provides part of the suppression for 
$b/t$, and allows $\tan\beta$ to be moderate ($\approx 15$) rather 
than large ($\approx 50$).

Here we make one brief comment. The 6D theory has a very rich structure 
with several different ways of assigning Higgs and matter multiplets 
leading to the same 5D $SU(5)$ theory. These different assignments can 
lead to a differing pattern of Yukawa couplings, reflecting the higher 
symmetries of the 6D theory.  For example, the Higgs doublets may be 
components of a ${\bf 10}$-plet of $SO(10)$ residing on the $x^6 = 0$ 
fixed line, rather than two ${\bf 5}$-plets of $SU(5)$ on the 
$x^6 = \pi R_6$ fixed line.  In this case, the Higgs zero modes 
fill out a single representation $({\bf 1}, {\bf 2}, {\bf 2})$ 
of $SU(4)_C \times SU(2)_L \times SU(2)_R$.  Thus, if the third 
generation matter is located on the $(x^5,x^6)=(0,0)$ fixed point as 
a ${\bf 16}$-plet of $SO(10)$, there will be a single Yukawa coupling 
at this location, giving rise to a unification of the $t$, $b$ and 
$\tau$ Yukawa couplings.

A $U(1)_R$ symmetry can be introduced in our theory, with the 
charge assignments given in Table~\ref{table:U1R}.  This $U(1)_R$ 
symmetry forbids all unwanted brane operators, such as 
$[H \bar{H}]_{\theta^2}$ and $[\hat{T}\hat{T}\hat{T}\hat{F}]_{\theta^2}$ 
on fixed points, and thus provides a complete solution to the 
doublet-triplet and proton decay problems \cite{Hall:2001pg, Hall:2001xb}. 
After supersymmetry is broken, $U(1)_R$ is (spontaneously) broken to the 
$Z_2$ subgroup, giving the usual $R$ parity of the MSSM.
\begin{table}
\begin{center}
\begin{tabular}{|c|c|cccc|cccc|cc|}  \hline 
  & $\theta^\alpha$ & $V$ & $\Sigma_5$ & $\Sigma_6$ & $\Phi$ 
  & $H$ & $H^c$ & $\bar{H}$ & $\bar{H}^c$ & $M$ & $M^c$ \\ \hline
  $U(1)_R$ & 1 & 0 & 0 & 0 & 2 & 0 & 2 & 0 & 2 & 1 & 1 \\ \hline
\end{tabular}
\end{center}
\caption{$U(1)_R$ charges for 4D vector and chiral superfields, 
normalized such that the superspace coordinates $\theta^\alpha$ of 
the 4D $N=1$ supersymmetry carry a unit charge.
Here, $M$ represents all the matter fields; for the example of the 
matter configuration in Fig.~\ref{fig:matter-config}, $M$ stands 
for $T_3$, $T_{1,2}$, $T'_{1,2}$ and $\Psi_{1,2,3}$, and $M^c$ for 
$T_{1,2}^c$, $T'^c_{1,2}$ and $\Psi_{1,2,3}^c$.}
\label{table:U1R}
\end{table}

Below the scale $M_6$, our theory is reduced to the 5D $SU(5)$ 
theory of Ref.~\cite{Hall:2001xb}, with a gauge interaction 
$U(1)_X \subset SO(10)/SU(5)$.  The breaking of $U(1)_X$ and the 
generation of small neutrino masses and a weak scale mass term ($\mu$ 
term) for the Higgs doublets can be accomplished along the lines 
of Ref.~\cite{Hall:2002ci}: neutrino masses are generated by the 
see-saw mechanism \cite{Seesaw} and the $\mu$ term is generated 
by the vacuum readjustment mechanism \cite{Hall:2002up}. All the 
positive features of the 5D $SU(5)$ theory are maintained; in 
particular, the successful predictions from gauge and Yukawa coupling 
unification are preserved, which we will see in more detail in 
section \ref{sec:unif}.

We finally discuss the other possibilities.  We can also construct 
realistic theories based on $G=SU(6)$, instead of $G=SO(10)$.
The boundary conditions for the bulk gauge supermultiplet are given 
by Eqs.~(\ref{eq:bc-Z}, \ref{eq:bc-T}), but now $P_5$ and $P_6$ are 
acting on the $SU(6)$ space.  To obtain the theory which reduces to 
5D $SU(5)$ below $M_6$, we take $P_6 = {\rm diag}(1,1,1,1,1,-1)$, 
giving $G_2 = SU(5) \times U(1)_X$.  For $P_5$, we have two choices 
$P_5 = {\rm diag}(1,1,1,-1,-1,-1)$ and $P_5 = 
{\rm diag}(-1,-1,1,1,1,1)$, giving $G_1 = SU(3) \times SU(3) \times U(1)$ 
and $G_1 = SU(4) \times SU(2) \times U(1)$, respectively.  In either 
case, $G_3 = SU(3)_C \times SU(2)_L \times U(1)_Y \times U(1)_X$, and 
the unbroken gauge symmetries below $M_5 = 1/R_5$ are the standard model 
gauge group with an extra $U(1)_X$.  The two Higgs hypermultiplets 
are introduced on the $x^6 = \pi R_6$ fixed line with quantum 
numbers given by $\{ H, H^c \}({\bf 5}, -2)$ and 
$\{ \bar{H}, \bar{H}^c \}(\bar{\bf 5}, 2)$.  The boundary conditions 
are given by Eq.~(\ref{eq:bc-hyper-5}) with $\eta_H = \eta_{\bar{H}} 
= -1$, giving two MSSM Higgs doublets at low energies.  The quarks 
and leptons are introduced either on the $SU(5) \times U(1)_X$ fixed 
point, $(x^5,x^6)=(0,\pi R_6)$, or fixed line, $x^6 = \pi R_6$, 
with Yukawa couplings located on the $(x^5,x^6)=(0,\pi R_6)$ fixed 
point. Again, the $U(1)_R$ symmetry can be introduced with the charge 
assignments given in Table~\ref{table:U1R}.

\subsection{Gauge-Higgs unification on asymmetric $T^2/(Z_2 \times Z'_2)$}
\label{subsec:higgs-gauge}

In this sub-section we construct theories where the Higgs and gauge 
fields are unified into a single 6D gauge supermultiplet. This type 
of theories has been considered in Ref.~\cite{Hall:2001zb}, but here 
we impose slightly different boundary conditions so that the theory 
is reduced to the 5D $SU(5)$ theory below the scale of the sixth 
dimension, $M_6$.

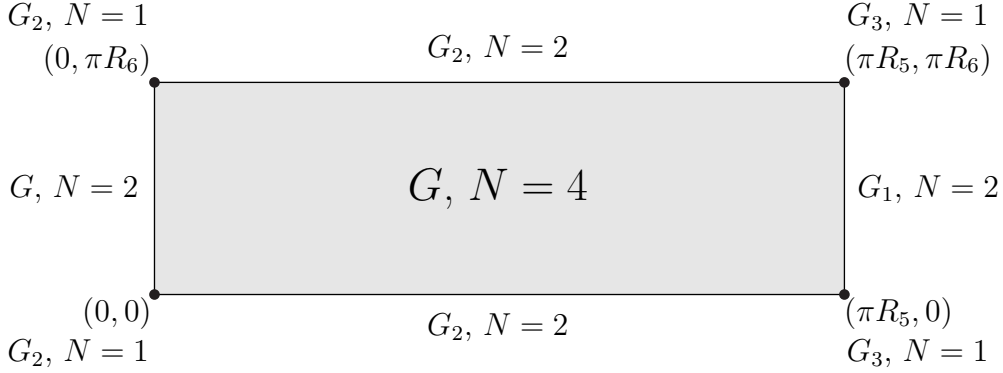
\begin{figure}
\begin{center}
\begin{picture}(300,130)(0,-5)
  \GBox(20,20)(280,100){0.9}
  \Text(150,60)[c]{\Large $G$, $N=4$}
  \Vertex(20,20){2}   \Text(19,19)[tr]{$(0,0)$} 
    \Text(18,3)[tr]{$G_2$, $N=1$}
  \Vertex(280,20){2}  \Text(281,19)[tl]{$(\pi R_5,0)$} 
    \Text(282,3)[tl]{$G_3$, $N=1$}
  \Vertex(20,100){2}  \Text(19,106)[br]{$(0,\pi R_6)$} 
    \Text(18,122)[br]{$G_2$, $N=1$}
  \Vertex(280,100){2} \Text(281,106)[bl]{$(\pi R_5,\pi R_6)$} 
    \Text(282,122)[bl]{$G_3$, $N=1$}
  \Text(14,60)[r]{$G$, $N=2$}  \Text(286,60)[l]{$G_1$, $N=2$}
  \Text(150,13)[t]{$G_2$, $N=2$} \Text(150,110)[b]{$G_2$, $N=2$}
\end{picture}
\caption{The structure of gauge and supersymmetries in the 2D extra 
dimensions with boundary conditions Eqs.~(\ref{eq:bc-Z-GH}, 
\ref{eq:bc-T-GH}).}
\label{fig:higgs-bulk}
\end{center}
\end{figure}

We again consider 6D $N=2$ supersymmetric gauge theories with gauge 
group $G$, compactified on the asymmetric $T^2/(Z_2 \times Z'_2)$ 
orbifold with $R_5 \gg R_6$.  The boundary conditions for the gauge 
multiplet are given by 
\begin{equation}
  \begin{array}{ccccc}
    V(x^5,x^6)        &=& P\, V(-x^5,x^6)\, P^{-1} 
      &=& P\, V(x^5,-x^6)\, P^{-1},         \\
    \Sigma_5(x^5,x^6) &=& -P\, \Sigma_5(-x^5,x^6)\, P^{-1} 
      &=& P\, \Sigma_5(x^5,-x^6)\, P^{-1},  \\
    \Sigma_6(x^5,x^6) &=& P\, \Sigma_6(-x^5,x^6)\, P^{-1}  
      &=& -P\, \Sigma_6(x^5,-x^6)\, P^{-1}, \\
    \Phi(x^5,x^6)     &=& -P\, \Phi(-x^5,x^6)\, P^{-1} 
      &=& -P\, \Phi(x^5,-x^6)\, P^{-1},
  \end{array}
\label{eq:bc-Z-GH}
\end{equation}
and
\begin{equation}
  \begin{array}{ccccc}
    V(x^5,x^6)        &=& P_5\, V(x^5+2\pi R_5,x^6)\, P_5^{-1} 
      &=& V(x^5,x^6+2\pi R_6), \\
    \Sigma_5(x^5,x^6) &=& P_5\, \Sigma_5(x^5+2\pi R_5,x^6)\, P_5^{-1} 
      &=& \Sigma_5(x^5,x^6+2\pi R_6), \\
    \Sigma_6(x^5,x^6) &=& P_5\, \Sigma_6(x^5+2\pi R_5,x^6)\, P_5^{-1} 
      &=& \Sigma_6(x^5,x^6+2\pi R_6), \\
    \Phi(x^5,x^6)     &=& P_5\, \Phi(x^5+2\pi R_5,x^6)\, P_5^{-1} 
      &=& \Phi(x^5,x^6+2\pi R_6),
  \end{array}
\label{eq:bc-T-GH}
\end{equation}
where $P$ and $P_5$ are matrices acting on the gauge space. The 
resulting gauge and supersymmetry structure in the 2D extra dimensions 
is summarized in Fig.~\ref{fig:higgs-bulk}.  Unlike the case of the 
previous section, the gauge symmetry structure is symmetric about 
$x^6 = \pi R_6/2$ in the sixth dimension while it is still asymmetric 
about $x^5 = \pi R_5/2$ in the fifth dimension.

An important point for the boundary conditions in 
Eqs.~(\ref{eq:bc-Z-GH}, \ref{eq:bc-T-GH}) is that they can leave low 
energy fields other than those from the 4D vector superfield $V$. 
Suppose we take $G=SU(6)$ and $P = {\rm diag}(1,1,1,1,1,-1)$.
Then, in the effective 5D theory below $M_6$, we have components of 
$\{ \Phi, \Sigma_6 \}$ propagating in the 5D bulk, in addition to the 
$SU(5) \times U(1)_X$ gauge multiplet coming from $\{ V, \Sigma_5 \}$. 
Since the components of $\{ \Phi, \Sigma_6 \}$ remaining below $M_6$ 
appear as 5D hypermultiplets transforming as $({\bf 5},-2) + 
(\bar{\bf 5},2)$ under $SU(5) \times U(1)_X$, we can identify these 
fields to be the two Higgs hypermultiplets located in the bulk of the 
effective 5D $SU(5) \times U(1)_X$ theory.  This allows us to construct 
theories where the gauge and bulk Higgs multiplets in 5D are unified 
into a single gauge multiplet in 6D.

Now, we explicitly construct a completely realistic theory.
We take $G=SU(6)$ and $P = {\rm diag}(1,1,1,1,1,-1)$ to reproduce 
two Higgs hypermultiplets, $\{ H, H^c \}$ and $\{ \bar{H}, 
\bar{H}^c \}$, in the effective 5D theory below $M_6$.
They arise from the $SU(6)/(SU(5) \times U(1)_X)$ components of 
the $\Phi$ and $\Sigma_6$ fields as $H+\bar{H} \subset \Phi$ and 
$H^c+\bar{H}^c \subset \Sigma_6$.  The matrix $P_5$ must be chosen 
as $P_5 = {\rm diag}(1,1,1,-1,-1,-1)$ to obtain the two MSSM Higgs 
doublets at low energies. This fixes the gauge structure to be 
$G_1 = SU(3) \times SU(3) \times U(1)$, $G_2 = SU(5) \times U(1)_X$ 
and $G_3 = SU(3)_C \times SU(2)_L \times U(1)_Y \times U(1)_X$.
The quark and lepton fields are introduced either on the $SU(5) \times 
U(1)_X$ fixed points, $(x^5,x^6)=(0,0),(0,\pi R_6)$, or fixed lines, 
$x^6=0,\pi R_6$.  If we locate matter fields in both $x^6 = 0$ and 
$x^6 = \pi R_6$ planes, the fields on the $x^6 = 0$ plane and on 
the $x^6 = \pi R_6$ plane do not have Yukawa couplings, giving texture 
zeros in the low energy 4D theory.  While it is interesting to work 
out more complicated cases, here we just present an example of 
realistic matter configuration which reduces to that of 
Ref.~\cite{Hall:2002ci} in the 5D effective theory below $M_6$.  
We introduce four chiral superfields $T_3({\bf 10}, 1) + 
F_{1,2,3}(\bar{\bf 5}, -3)$ on the $(x^5,x^6)=(0,\pi R_6)$ fixed 
point, and four hypermultiplets $\{ T_{1,2} + T_{1,2}^c \}({\bf 10}, 1) 
+ \{ T'_{1,2} + T'^c_{1,2} \}({\bf 10}, 1)$ with $\eta_{T_{1,2}} = 
-\eta_{T'_{1,2}} = 1$ on the $x^6 = \pi R_6$ fixed line.\footnote{
As in the previous sub-section, we have made choices for the $U(1)_X$ 
quantum numbers of matter so that Yukawa couplings with the Higgs 
in $\Phi$ are allowed.}
We also introduce three right-handed neutrino fields to cancel 
$U(1)_X$ anomalies, which can be located either on the fixed point as 
$N({\bf 1}, 5)$ or on the fixed line as $\{ N + N^c \}({\bf 1}, 5)$ 
with $\eta_{N} = 1$. This completes the three generations of matter 
(including right-handed neutrinos) at low energies.  

The Yukawa couplings of the form in Eq.~(\ref{eq:yukawa}) are 
introduced on the $(x^5,x^6)=(0,\pi R_6)$ fixed point, but now $H$ and 
$\bar{H}$ fields arise from the component, $\Phi$, of the 6D gauge 
multiplet.  It is important to notice that $\Phi$ transforms linearly 
under the gauge group so that it can have Yukawa couplings to quarks 
and leptons without contradicting to the higher dimensional gauge 
invariance \cite{Hall:2001zb}. Below $M_5$, we obtain the 4D Yukawa 
couplings of the form Eq.~(\ref{eq:4D-yukawa}) with $\epsilon_6=1$, 
which explains a part of the observed structure of fermion mass 
matrices, including the unified relation for $m_b/m_\tau$, the absence 
of the corresponding relation for $m_s/m_\mu$, large neutrino mixing 
angles, and a hierarchy of masses between the third and the first 
two generations. 

In the theories with gauge-Higgs unification, the $R$ charges of 
the Higgs fields are determined because they are part of the gauge 
multiplet.  In particular the Higgs fields, $H$ and $\bar{H}$, must 
transform non-trivially under the $R$ symmetry so that the 6D 
supersymmetric gauge kinetic term is invariant.  This forces us to 
use a discrete $R$ symmetry to forbid unwanted brane operators while 
keeping the Yukawa couplings. We thus consider the $Z_{4,R}$ symmetry 
with charge assignments given in Table~\ref{table:Z4R}.  We require 
that all the operators in the Lagrangian must be invariant under 
$Z_{4,R}$: the terms in the superpotential (K\"{a}hler potential) 
must carry charges of $+2$ ($0$) mod 4.  This forbids all unwanted 
brane operators, such as $[H\bar{H}]_{\theta^2}$ (i.e. 
$[\Phi^2]_{\theta^2}$) and $[\hat{T}\hat{T}\hat{T}\hat{F}]_{\theta^2}$, 
completing the solutions to the doublet-triplet and proton decay 
problems.
\begin{table}
\begin{center}
\begin{tabular}{|c|c|cccc|cc|}  \hline 
  & $\theta^\alpha$ & $V$ & $\Sigma_5$ & $\Sigma_6$ & $\Phi$ 
  & $M$ & $M^c$ \\ \hline
  $Z_{4,R}$ & 1 & 0 & 0 & 0 & 2 & 0 & 2 \\ \hline
\end{tabular}
\end{center}
\caption{$Z_{4,R}$ charges for 4D vector and chiral superfields, 
normalized such that the superspace coordinates $\theta^\alpha$ of 
the 4D $N=1$ supersymmetry carry a unit charge.  Here, $M$ 
collectively represents all the matter fields.  All the operators 
in the Lagrangian must be invariant under $Z_{4,R}$ (i.e. must have 
vanishing charges mod $4$.)}
\label{table:Z4R}
\end{table}

The $U(1)_X$ breaking and the generation of small neutrino masses 
can be achieved preserving $Z_{4,R}$ symmetry.  Specifically, we 
introduce $[X(B\bar{B} - \Lambda^2) + \bar{B}NN]_{\theta^2}$ on the 
$(x^5,x^6) = (0,\pi R_6)$ brane, where $X({\bf 1},0), B({\bf 1},10)$ 
and $\bar{B}({\bf 1},-10)$ are chiral superfields with the $SU(5) 
\times U(1)_X$ quantum numbers given in the parentheses; the $Z_{4,R}$ 
charges for the $X$, $B$ and $\bar{B}$ fields are given by $2$, $-2$ 
and $2$, respectively.  This brane superpotential gives vacuum 
expectation values $\langle B \rangle = \langle \bar{B} \rangle = 
\Lambda$ and consequently the right-handed neutrino masses of order 
$\Lambda$. Taking $\Lambda \simeq 10^{14}~{\rm GeV}$, we obtain small 
neutrino masses of desirable sizes through the see-saw mechanism. The 
expectation values $\langle B \rangle = \langle \bar{B} \rangle \neq 0$ 
break both $Z_{4,R}$ and $U(1)_X$ symmetries, but it leaves another 
unbroken discrete $Z'_{4,R}$ symmetry that is a linear combination 
of $Z_{4,R}$ and $U(1)_X$: $Z'_{4,R} = Z_{4,R} + (1/5) U(1)_X$.
(To make all charges integer, we have to take a linear combination, 
 $Z_{4,R} + (1/5) U(1)_X + (24/5) U(1)_Y$.)  This $Z'_{4,R}$ symmetry 
is sufficient to forbid all the unwanted operators, and thus no large 
$\mu$ term or dimension four/five proton decay operators are generated 
by this symmetry breaking.  A $\mu$ term of the order of the weak 
scale can be generated though the $Z'_{4,R}$ breaking encoded in 
supersymmetry breaking parameters, through the mechanism of 
Ref.~\cite{Hall:2002up}.  The usual MSSM $R$ parity, which is the 
subgroup of $Z'_{4,R}$, could remain unbroken.

\section{Gauge and Yukawa Coupling Unification}
\label{sec:unif}

In this section we show that our asymmetric 6D theories preserve 
successful predictions from gauge and Yukawa couplings obtained in 
Refs.~\cite{Hall:2001xb, Hall:2002ci}.  We present an explicit 
analysis in the present theories, following the general analysis of 
Ref.~\cite{Hall:2001xb}, which elucidates some of the general features 
for the behavior of gauge couplings in higher dimensional unified 
field theories.

We first consider the effective (Wilsonian) action at the cutoff scale 
$M_s$.  No matter what the physics above $M_s$ is, the general form for 
the gauge kinetic terms are given by
\begin{eqnarray}
  S &=& \int d^6x \; \biggl[ \frac{1}{g_6^2} F_{\mu\nu}F_{\mu\nu} 
\nonumber\\
 && + \sum_{\hat{x}_5=0,\pi R_5} \delta(x_5-\hat{x}_5)
      \frac{1}{g_{5,(\hat{x}_5,*)}^2} F_{\mu\nu}F_{\mu\nu} 
    + \sum_{\hat{x}_6=0,\pi R_6} \delta(x_6-\hat{x}_6) 
      \frac{1}{g_{5,(*,\hat{x}_6)}^2} F_{\mu\nu}F_{\mu\nu}
\nonumber\\
 && + \sum_{\hat{x}_5=0,\pi R_5} \sum_{\hat{x}_6=0,\pi R_6} 
      \delta(x_5-\hat{x}_5) \delta(x_6-\hat{x}_6)
      \frac{1}{g_{4,(\hat{x}_5,\hat{x}_6)}^2} F_{\mu\nu}F_{\mu\nu} 
      \biggr].
\label{eq:gen-gke}
\end{eqnarray}
Here, a term located on a subspace only respects the gauge symmetry 
operative on that subspace. Note that this form is guaranteed by the 
restricted unified gauge symmetry (position dependent gauge symmetry) 
of the effective higher dimensional field theory below $M_s$.  The 4D 
gauge couplings for the zero modes are obtained by integrating over 
the extra dimensions:
\begin{equation}
  \frac{1}{g_{4D,i}^2} = \frac{\pi^2 R_5 R_6}{g_6^2} 
    + \sum_{\hat{x}_5=0,\pi R_5} \frac{\pi R_6}{g_{5,(\hat{x}_5,*)}^2} 
    + \sum_{\hat{x}_6=0,\pi R_6} \frac{\pi R_5}{g_{5,(*,\hat{x}_6)}^2} 
    + \sum_{\hat{x}_5=0,\pi R_5} \sum_{\hat{x}_6=0,\pi R_6} 
      \frac{1}{g_{4,(\hat{x}_5,\hat{x}_6)}^2},
\label{eq:4D-gc-1}
\end{equation}
where $i$ runs for $SU(3)_C$, $SU(2)_L$, $U(1)_Y$ and $U(1)_X$.
Since the theory is strongly coupled at the scale $M_s$, we can estimate 
the coefficients for various operators by requiring that all loop 
contributions become comparable at this scale.  By carefully evaluating 
loop expansion parameters, we find that $1/g_6^2 \simeq C M_s^2/16\pi^4$, 
$1/g_{5,(\hat{x}_5,*)}^2 \simeq 1/g_{5,(*,\hat{x}_6)}^2 \simeq 
C M_s/16\pi^3$ and $1/g_{4,(\hat{x}_5,\hat{x}_6)}^2 \simeq C/16\pi^2$, 
giving 
\begin{eqnarray}
  \frac{1}{g_{4D,i}^2} &\simeq& \frac{C(M_s R_5)(M_s R_6)}{16 \pi^2} 
    + \frac{C(M_s R_6)}{16\pi^2} + \frac{C(M_s R_5)}{16\pi^2} 
    + \frac{C}{16\pi^2}
\nonumber\\
  &=& \frac{C(M_s R_5)(M_s R_6)}{16 \pi^2} \biggl( 
    1 + \frac{1}{M_s R_5} + \frac{1}{M_s R_6} 
    + \frac{1}{(M_s R_5)(M_s R_6)} \biggr),
\label{eq:4D-gc-2}
\end{eqnarray}
where $C$ is a group theoretical factor, and the first, second, third 
and fourth terms represent contributions from the 6D bulk, 5D short 
fixed lines, 5D long fixed lines and 4D fixed points, respectively;
here we have omitted unknown order-one coefficients for each term.
We thus find that the contributions from short fixed lines and 
fixed points are suppressed by the large volume factor $M_s R_5$, 
while those from long fixed lines have only small suppression by 
$M_s R_6$.  (Recall, $M_s R_5 \gg M_s R_6 \sim 1$.)
However, since the gauge symmetries on the long fixed lines (and in the 
bulk) contain $SU(5)$, this is sufficient for guaranteeing successful 
gauge coupling unification.  In other words, although there are unknown 
$SU(5)$-violating contributions to the 4D (zero-mode) gauge couplings 
at $M_s$ coming from the gauge kinetic terms localized on 5D short fixed 
lines and 4D fixed points, they are suppressed by the large radius of 
the fifth dimension.  Here we take $M_s R_5 \simeq 30$ to suppress these 
unknown contributions to a negligible level.  This implies that $M_s R_6$ 
must be a factor of a few to obtain order-one 4D gauge coupling constants, 
$g_{4D,i} =  O(1)$.

Having obtained gauge coupling unification at the scale $M_s$, we next 
consider radiative corrections coming from an energy interval between 
$M_s$ and $M_6$.  In this energy interval, 4D gauge couplings receive 
power corrections.  From the higher dimensional point of view, these 
corrections arise from radiative corrections to the gauge kinetic terms 
localized on 5D fixed lines.  (Since we have 6D $N=2$ supersymmetry in 
the bulk, the bulk gauge kinetic term does not receive any quantum 
correction.)  However, we can show that the size of the power 
corrections is always smaller than the tree-level estimates given in 
Eq.~(\ref{eq:4D-gc-2}).  Specifically, the $SU(5)$-violating power-law 
corrections to $1/g_{4D,i}^2$ are given by $\simeq (b/16\pi^2)(M_s/M_6)$, 
where $b$ is an appropriate beta-function coefficients, so that these 
corrections are at most the same size with the tree-level values given 
at $M_s$.\footnote{
In addition, there are $SU(5)$-symmetric power-law corrections of 
size $\simeq (b/16\pi^2)(M_s/M_5)$, which also do not exceed the 
tree-level values at $M_s$.}
Note that this is a general consequence of the effective field theory 
framework. In an effective field theory, power corrections are scheme 
dependent and can always be absorbed into the definitions of the 
tree-level parameters at the cutoff scale.  Therefore, by appropriately 
estimating the size of the tree-level terms at the cutoff scale, we 
can always forget about the presence of power corrections; these are 
contributions coming from the physics at or above $M_s$ and cannot be 
computed in the effective field theory framework.  Unlike power-law 
corrections, logarithmically divergent contributions and finite 
contributions are calculable and meaningful quantities in the 
effective field theory.  In our case, there are logarithmic 
contributions to $1/g_{4D,i}^2$ coming from the running of the gauge 
kinetic operators localized on the 4D fixed points, whose sizes are 
given by $\simeq (b'/16\pi^2)\ln(\pi M_s/M_6)$.  However, since 
$\ln(\pi M_s/M_6)$ is not a large quantity, they are not much larger than 
unknown tree-level contributions of order $C/16\pi^2$.  Therefore, here 
we do not include this contribution to the calculation of $\alpha_s(M_Z)$. 
We also neglect the finite correction at the scale $M_6$, since it is 
also similar in size to unknown tree-level contributions.  However, it is 
important to notice that these are calculable contributions and, 
if one wants, can be included for the prediction of $\alpha_s(M_Z)$.

Now, we match our 6D theory to the effective 5D theory.  Since we 
have found that quantum corrections between $M_s$ and $M_6$ do not 
change the tree-level coefficients much, the gauge kinetic terms at the 
scale $M_6$ are still given by Eq.~(\ref{eq:gen-gke}) with various 
coefficients taking the sizes given just above Eq.~(\ref{eq:4D-gc-2}).
Integrating over the sixth dimension, we obtain the effective action 
at $M_6$ in the effective 5D theory:
\begin{equation}
  S = \int d^5x \; \biggl[ \frac{1}{g_5^2} F_{\mu\nu}F_{\mu\nu} 
  + \sum_{\hat{x}_5=0,\pi R_5} \delta(x_5-\hat{x}_5)
    \frac{1}{g_{4,\hat{x}_5}^2} F_{\mu\nu}F_{\mu\nu} \biggr],
\label{eq:gen-gke-5d}
\end{equation}
where various coefficients are given by
\begin{equation}
  \frac{1}{g_5^2} 
    = \frac{\pi R_6}{g_6^2} 
      + \sum_{\hat{x}_6=0,\pi R_6} \frac{1}{g_{5,(*,\hat{x}_6)}^2}
    \simeq \frac{C M_s^2 R_6}{16 \pi^3} 
      \biggl( 1 + \frac{1}{M_s R_6} \biggr),
\label{eq:5D-gcs-1}
\end{equation}
\begin{equation}
  \frac{1}{g_{4,\hat{x}_5}^2} 
    = \frac{\pi R_6}{g_{5,(\hat{x}_5,*)}^2} 
      + \sum_{\hat{x}_6=0,\pi R_6} \frac{1}{g_{4,(\hat{x}_5,\hat{x}_6)}^2}
    \simeq \frac{C M_s R_6}{16 \pi^2}
      \biggl( 1 + \frac{1}{M_s R_6} \biggr).
\label{eq:5D-gcs-2}
\end{equation}
Thus the gauge structure of the bulk (brane-localized) kinetic energy, 
$g_5$ ($g_{4,\hat{x}_5}$), arises from those of the bulk and long 
fixed lines (short fixed lines and fixed points) in the original 6D 
theory.

We first consider the bulk gauge coupling, $g_5$.  It comes from a 
linear combination of the couplings of the 6D bulk and 5D long fixed 
lines in the original theory.  In any of the models discussed in the 
previous section, there is a 5D long fixed line on which the gauge 
symmetry is only $SU(5) \times U(1)_X$.  Since the effect from 5D long 
fixed lines on $g_5$ is unsuppressed (suppressed only by a small volume 
factor of $M_s R_6$), we do not find any particular relation between 
the bulk gauge couplings for $SU(5)$ and $U(1)_X$ in the effective 
5D theory.  On the other hand, $g_5$ clearly respects $SU(5)$, since 
the $SU(5)$ gauge symmetry remains unbroken in the bulk of the 5D 
effective theory; in fact, the bulk and long fixed lines of the original 
theory always respect $SU(5)$ by construction.  Similar considerations 
show that the brane-localized gauge couplings $g_{4,\hat{x}_5=0}$ and 
$g_{4,\hat{x}_5=\pi R}$ respect (only) $SU(5) \times U(1)_X$ and 
$SU(3)_C \times SU(2)_L \times U(1)_Y \times U(1)_X$, respectively.

Summarizing so far, we have obtained the effective 5D $SU(5) \times 
U(1)_X$ theory at $M_6$, where the sizes of the bulk and brane gauge 
couplings are given by $1/g_5^2 \simeq C M_s^2 R_6/16 \pi^3$ and 
$1/g_{4,\hat{x}_5}^2 \simeq C M_s R_6/16 \pi^2$, respectively.  The 
bulk gauge coupling and the brane gauge coupling at $x^5=0$ respect 
$SU(5) \times U(1)_X$ symmetry, while the brane coupling at $x^5=\pi R_5$ 
respects only $SU(3)_C \times SU(2)_L \times U(1)_Y \times U(1)_X$.
Since $M_s R_6$ is not much larger than unity, we find that the situation 
is almost the same with the minimal 5D theory in Ref.~\cite{Hall:2001xb} 
with the cutoff scale replaced by $M_6$.  Of course, now the bulk and 
brane gauge couplings are slightly smaller than the case of a single 
extra dimension, $1/g_5^2 \simeq C M_s/16 \pi^3$ and $1/g_{4,\hat{x}_5}^2 
\simeq C/16 \pi^2$, due to the volume suppression from the sixth 
dimension, leading to a somewhat smaller size for the fifth dimension. 
However, this effect is numerically not so large that the correction to 
gauge coupling unification is still dominated by a logarithmic 
contribution coming from the energy interval between $M_6$ and $M_5$. 
Below we explicitly show that this contribution corrects the prediction 
for $\alpha_s(M_Z)$ to the values which agree well with experiment, by 
looking at the behavior of the gauge couplings at the energy scale 
below $M_6$.

The bulk and brane gauge couplings receive power-law and logarithmic 
corrections, respectively, in the energy interval between $M_6$ and 
$M_5$. The correction to the bulk coupling, $1/g_5^2$, is $SU(5)$ 
symmetric and has a size $\simeq (b/16\pi^2)(M_6/M_5)$. It is dominated 
at the scale $M_6$ and, in fact, can also be interpreted as the finite 
correction at $M_6$.  This correction does not contribute to the 
prediction of $\alpha_s(M_Z)$ because it is $SU(5)$ symmetric. On the 
other hand, the brane couplings receive logarithmic contributions, 
which intrinsically arise from the physics between $M_6$ and $M_5$ and 
cannot be attributed to any other corrections.  Furthermore, since the 
fixed point at $x^5 = \pi R_5$ does not respect $SU(5)$, they are not 
$SU(5)$ symmetric and contribute to the prediction of $\alpha_s(M_Z)$. 
The beta-function coefficients for this logarithmic evolution are given 
by $(b_1,b_2,b_3) = (0,-4,-6)$ plus an $SU(5)$ symmetric piece coming 
from matter \cite{Hall:2001pg}, so that we obtain
\begin{equation}
  \delta\alpha_s 
    \simeq -\frac{3}{7\pi} \alpha_s^2 
    \left( \ln\frac{\pi M_6}{M_5} + \Delta \right),
\label{eq:as-final}
\end{equation}
where $\Delta = O(1)$ represents effects from unknown brane-localized 
operators at $M_s$, logarithmic corrections between $M_s$ and $M_6$, 
and finite corrections at $M_6$ and $M_5$ \cite{Hall:2001xb}. 
Substituting $M_6/M_5 \simeq 20$ as an example, we obtain 
$\delta\alpha_s \simeq -0.01$.  Although the error is larger than that 
in the case where the theory is five dimensional up to the cutoff scale, 
the correction in Eq.~(\ref{eq:as-final}) still significantly improves 
the agreement with data.  Therefore, we find that our theories retain 
the successful feature of the two-stage gauge coupling unification of 
the minimal 5D $SU(5)$ theory, with $M_5 \approx 10^{15}~{\rm GeV}$ 
and with $M_6$ and $M_s$ in the region of $10^{16}-10^{17}~{\rm GeV}$.

We next consider Yukawa coupling unification.  In our theories, 
whether the Yukawa couplings are unified or not depends on the location 
of matter. Here we consider the case where the third generation matter 
is located on the $x^5 = 0$ subspace (either on the fixed line or 
on a fixed point) so that we have $b/\tau$ Yukawa unification.
The behavior of Yukawa couplings is quite different from that of 
gauge couplings.  In fact, by integrating out the physics above $M_5$, 
we find that no violation of $SU(5)$ is felt by bottom and tau Yukawa 
couplings above $M_5$.  Thus, they start to deviate at $M_5$, giving 
the correction to the 4D prediction for the bottom quark mass
\begin{equation}
  \frac{\delta m_b}{m_b} 
    \simeq - \frac{20g^2 - 5y_t^2}{112\pi^2} \ln\frac{\pi M_6}{M_5},
\label{eq:delta-mb-final}
\end{equation}
where $g$ and $y_t$ are the 4D gauge and Yukawa couplings around the 
scale $M_5$ \cite{Hall:2002ci}. Again, although this expression is not 
as precise as the case of the exact single extra dimension, we find that 
it improves the agreement with data.

So far, we have considered the case where there is a non-negligible 
energy interval in which the physics is described by perturbative 6D 
theories. Here we comment on the possibility of taking the limit $M_s R_6 
\rightarrow 1$.  In this limit, the predictions for gauge and Yukawa 
coupling unification are reduced to those of Refs.~\cite{Hall:2001xb, 
Hall:2002ci}, where the theory is five dimensional up to the cutoff scale. 
Although it does not make much sense to talk about the field theoretic 6D 
theories in this case, we expect that some aspects of our constructions, 
such as the patterns of gauge symmetry breaking and/or the unification 
of gauge and Higgs fields, persist even in this limit, presumably as 
intermediate steps for string compactification. It would be interesting 
to further pursue the present line of constructions to more symmetrical 
theories having a larger gauge group and/or number of dimensions, such 
as $E_8$ in 10D.

\section{Conclusions}
\label{sec:concl}

We have constructed maximally supersymmetric $SO(10)$ and $SU(6)$ 
models in 6D, on the orbifold $T^2/(Z_2 \times Z'_2)$, with one 
dimension, $R_5$, much larger than the other, $R_6$. A set of 
boundary conditions leads to a pattern of local breaking of the gauge
symmetry such that, at distance scales larger than $R_6$, the theory
reduces to $SU(5) \times U(1)_X$ in 5D. The $SU(5)$ sector corresponds 
to the highly predictive and successful minimal unified theory in 5D, 
while the $U(1)_X$ gauge symmetry allows an understanding of the scales 
of both neutrino masses and the $\mu$ parameter. Thus these highly 
symmetric $SO(10)$ and $SU(6)$ theories reproduce the predictions, 
for example for $\alpha_s(M_Z)$, $m_b/m_\tau$ and proton decay, of 
the minimal 5D $SU(5)$ theory \cite{Hall:2001xb, Hall:2002ci}, although 
with a slightly reduced precision. In the case that the theory is five 
dimensional up to the scale of strong coupling, we have previously 
argued that, with $F_i$ all on the $SU(5)$ invariant brane as expected 
from $b/\tau$ Yukawa unification and the observed large neutrino mixing 
angles, exchange of the broken $SU(5)$ gauge boson leads to 
$p \rightarrow l^+ \pi^0, l^+ K^0, \bar{\nu} \pi^+, \bar{\nu} K^+$ 
($l = e, \mu$) with a lifetime of order $10^{34}~{\rm years}$ 
\cite{Hall:2002ci}. In the case of 6D, with $M_s/M_6 \lsim 2$, the 
broken gauge boson mass is increased only by a factor $\lsim 3$, so 
that the proton decay lifetime is still expected to be in the range 
$10^{34} - 10^{36}~{\rm years}$.  Finally, the lightness of the Higgs 
doublets can be understood whether they originate as fields in 5D or, 
for the case of $SU(6)$, in 6D, where they are identified as components 
of the gauge supermultiplet. 

These more unified theories in 6D offer new possibilities for 
phenomenology beyond those of the 5D theory. The mass ratio $m_t/m_b$ 
may arise partially from $M_s R_6$, allowing moderate values for 
$\tan\beta$; or alternatively the $t,b$ and $\tau$ Yukawa couplings 
may all be unified in an $SO(10)$ theory, with $\tan\beta \approx 50$. 
In the 5D theory, fields located at the fixed point with $SU(3)_C 
\times SU(2)_L \times U(1)_Y$ gauge symmetry are not forced to have 
their hypercharge quantized. However, if the field propagates in 
a sixth dimension, even if it is very small, the hypercharge will be 
quantized as long as $U(1)_Y$ is embedded into a non-Abelian gauge 
group on this fixed line. Finally, a larger variety of locations for 
matter increases the number of ways in which texture zeros may occur 
in the Yukawa matrices.

\section*{Acknowledgments}

Y.N. thanks the Miller Institute for Basic Research in Science 
for financial support.  This work was supported in part by the Director, 
Office of Science, Office of High Energy and Nuclear Physics, of the U.S. 
Department of Energy under Contract DE-AC03-76SF00098, and in part 
by the National Science Foundation under grant PHY-00-98840.

\end{document}